\documentclass[prl,aps,twocolumn,showpacs]{revtex4-2}
\usepackage{color,graphicx,bm,amsmath}

\begin{document}

\newcommand{\half}{\ensuremath{\frac{1}{2}}}
\newcommand{\degree}{\ensuremath{^{\circ}}}
\newcommand{\adag}{\ensuremath{a^{\dagger}}}
\newcommand{\bra}[1]{\ensuremath{\langle #1 |}}
\newcommand{\ket}[1]{\ensuremath{ | #1 \rangle}}
\newcommand{\braket}[2]{\ensuremath{\langle #1 | #2 \rangle}}
\newcommand{\trace}[1]{\ensuremath{{\rm Tr}\{#1\}}}
\newcommand{\nbar}{\ensuremath{\overline{n}}}
\newcommand{\nav}{\ensuremath{\langle n \rangle}}
\newcommand{\gbar}{\ensuremath{\overline{\gamma}}}

\newcommand{\gf}[1]{\ensuremath{g^{(1)}(#1)}}
\newcommand{\gs}[1]{\ensuremath{g^{(2)}(#1)}}

\newcommand{\mattwo}[4]{
\left(
\begin{array}{cc}
 #1 & #2 \\
 #3 & #4
\end{array}
\right)
}

\newcommand{\vectwo}[2]{
\left(
\begin{array}{c}
 #1 \\
 #2 
\end{array}
\right)
}

\title{Topological Protection in Disordered  Photonic Multilayers and Transmission Lines}

\author{D. M. Whittaker}
\affiliation{Department of Physics and Astronomy, University of
Sheffield, Sheffield S3 7RH, UK.}

\author{R. Ellis}
\affiliation{Department of Physics and Astronomy, University of
Sheffield, Sheffield S3 7RH, UK.}

\date{\today}

\begin{abstract} 
The Su-Schrieffer-Heeger (SSH) model is the simplest
example of a lattice with non-trivial topology. It supports
mid-gap topologically protected states, whose energies are unaffected by
disorder. We show that photonic multilayer structures provide an exact
implementation of an SSH lattice, provided each layer has the same
propagation thickness. From this, it follows that the cavity mode in a
conventional semiconductor microcavity is a protected SSH mid-gap state. 
We demonstrate this experimentally using controlled disorder
in a mathematically equivalent system, a radio
frequency transmission line made from sections of coaxial cable with
high and low impedances.  We also show theoretically that transmission
lines connected to form networks map onto topologically interesting
lattices in in higher dimensions.  
\end{abstract}

\maketitle

\begin{figure}

\begin{center} \mbox{ \includegraphics[scale=0.2]{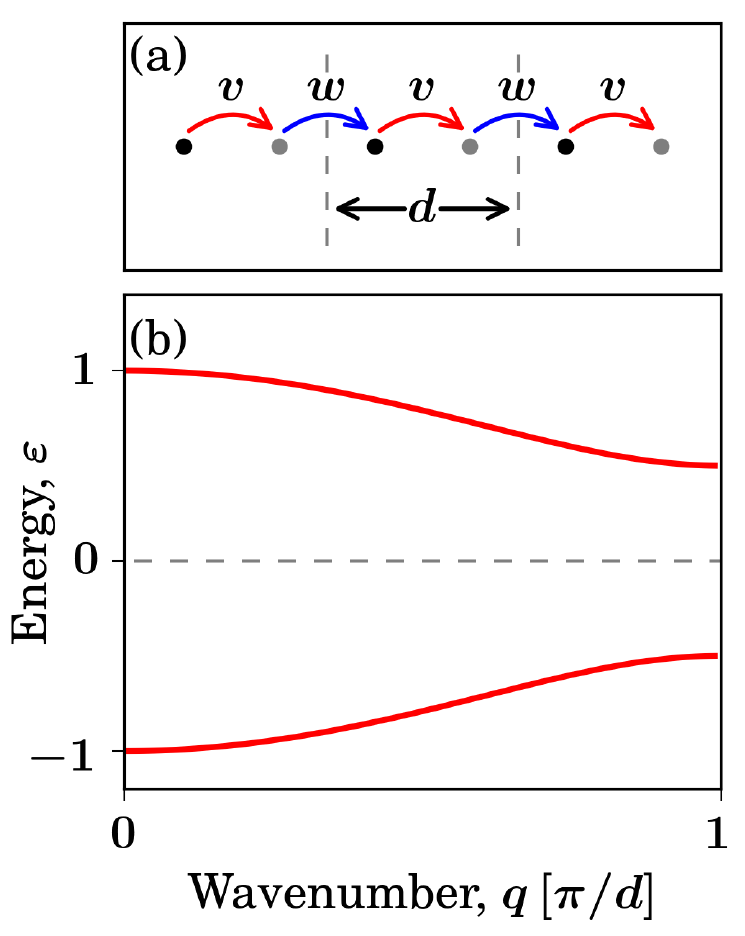} } \end{center}
\caption{(a) Schematic of part of an SSH lattice, showing the intra- and
inter-cell hopping amplitudes, $v$ and $w$. A single period, of length d,
consists of two sites. (b) Band structure for the case
$v=1/4$, $w=3/4$. The band structure is symmetric about the dashed line at
$\varepsilon=0$, the chiral point. } 
\label{fig-ssh}
\end{figure}

The Su-Schrieffer-Heeger (SSH) model\cite{SSH} is well known as the
`toy' system used to introduce the ideas of topology in solid state
physics. It consists of a one dimensional chain of sites coupled by
alternating weak and strong bonds.  The SSH  model has chiral
symmetry, meaning that its spectrum is exactly symmetric when
reflected about the middle of the band-gap. This symmetry is a
prerequisite for non-trivial topology in one dimension. It provides
topological protection for mid-gap states, so their energies are
unaffected by variations in the parameters of the system.

Although the SSH model has its origins in polymers, there have been
numerous attempts to construct
photonic\cite{photonic-ozawa,photonic-malkova,microwave-poli,polariton-st-jean,plasmon-bleckmann},
acoustic\cite{acoustic-li,acoustic-esmann} and cold-atom\cite{atom-atala} systems
in order to implement the topological physics it describes. They all
share the idea of creating a one-dimensional chain of resonators with
alternating strong and weak couplings. However, this is not in itself
enough to ensure chiral symmetry; it can be broken by effects such as
on-site potentials, second nearest neighbour hopping and coupling to
states not included in the manifold described by the model.  Such
effects are unavoidable in these complex systems, so they do
not exhibit exact chiral symmetry, and thus do not have the topology
of the SSH model. We show that, by contrast, there is a class of very
simple one-dimensional systems, described mathematically by transfer
matrices, which possess exact chiral symmetry and map perfectly onto SSH
chains. We consider two examples: multilayer photonic structures,
where the layers have different refractive indicies, and transmission
lines with sections of different impedances. These structures have
chiral symmetry if the propagation length of each layer, or line
section, is identical.

A consequence of this mapping is that in semiconductor microcavity
structures\cite{microcavity-weisbuch,microcavity-skolnick}, the cavity
mode is an SSH mid-gap state. Disorder can be introduced into these
structures, without breaking the chiral symmetry, by varying the
refractive indicies and layer thicknesses in a way which leaves the
propagation lengths unchanged.  They thus provide an ideal
experimental platform for demonstrating the topological protection
which is central to the SSH model. For the photonic case,
we show numerically that when disorder is introduced in this way, the
energy of the cavity mode is unaffected. We also
demonstrate the protection experimentally in a mathematically
analogous system of coaxial
cables\cite{coax-schneider,coax-munday,coax-sanchez-lopez}, where an
ensemble of disordered structures can be constructed very easily by
swapping in cables of different impedances.

It is important to distinguish the spectral chiral symmetry considered here
from the spatial inversion symmetry possessed by all periodic bilayer
structures.  Inversion symmetry has been demonstrated to have
topological consequences\cite{inversion-zak}, which have been discussed in
the context of photonic multilayers by Xiao et
al\cite{multilayer-xiao}, who show that it can be used to determine
the conditions for finding a localised mode at an interface. However,
this is not the topology of the SSH model. That requires chiral
symmetry, which can also be found in multilayers, but only if the
layers have the same propagation lengths. As we show, chiral symmetry
can be maintained in the presence of disorder, while inversion
symmetry is inevitably broken, so is not associated with topological protection.

The basic SSH model consists of a one dimensional chain of sites, all
with the same energy. A particle can hop between adjacent sites, with
hopping amplitudes alternating along the chain, as in
Fig.\ref{fig-ssh}(a). The system has
two energy bands, shown in Fig.1(b), and also two distinct
topologies, depending on which of the inter- and intra-cell hopping
amplitudes is greater.  An important property of the model is that the
spectrum possesses chiral symmetry; it is exactly symmetric when
reflected about the middle of the band-gap, which we will call the
chiral point. The chiral symmetry is not restricted to the simplest
case of alternating hopping amplitudes; it is present in any chain
where there are only nearest neighbour hoppings, and the on-site
energies are all the same. In such systems, energy states
generally occur in pairs, which map onto each other when reflected
about the chiral point. However, there are also special states, found
at the chiral point, which map onto themselves, and are thus unpaired.
In the SSH model, these are mid-gap states. They can occur at the end
of a chain, or at an interface between two domains of different
topology, created by repeating one of the hoppings in adjacent links.
When the system parameters are varied, perhaps due to disorder, the
energy of these states cannot change: if an unpaired state were to
move from the chiral point, it would violate the chiral symmetry. Such
states are said to be topologically protected.

\section*{Photonic Multilayers}

\begin{figure}

\begin{center} \mbox{ \includegraphics[scale=0.22]{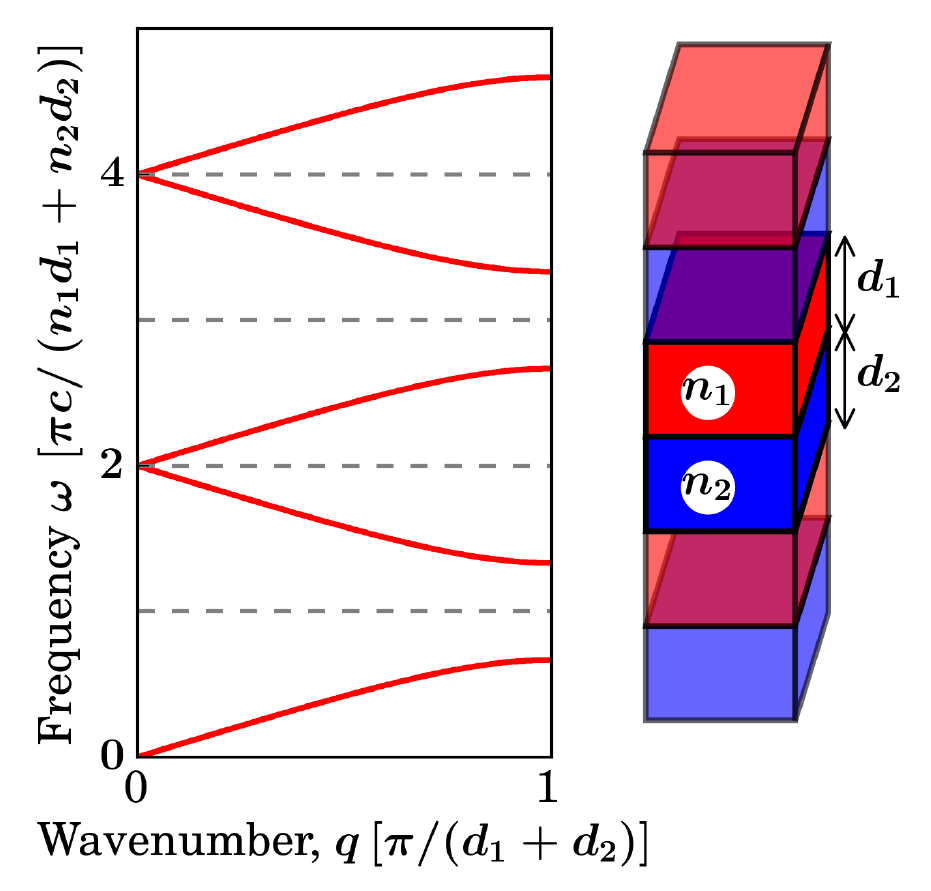} }\end{center} 
\caption{Band structure of a Bragg stack. The schematic illustrates a
section of the stack, which has a period of
two layers with refractive indicies $n_1=3$ and $n_2=1$. The layer
thicknesses satisfy the Bragg condition $n_1 d_1 = n_2 d_2$,
giving chiral symmetry. The band structure is periodic in
frequency, with period $2 \pi c /(n_1 d_1 + n_2 d_2)$. The dashed
lines show the chiral points; the bandstructure is exactly symmetric
when reflected about these lines. }
\label{fig-bragg} 
\end{figure}

The photonic structures considered here consist of a one dimensional
series of layers, characterised by a refractive index $n$ and
thickness $d$. The simplest such structure, the Bragg stack, is a
periodically repeated bilayer, which has the bandstructure shown in
Fig.\ref{fig-bragg}.  By introducing defects in the periodicity, such
as repeating layers, localised states can be created in the band gaps.

The simplest way to calculate the properties of such a layered
structures is to use the transfer matrix formalism.  The transfer
matrix for a layer relates the amplitudes of the electric and magnetic
fields, $E$ and $B$, on either side of the layer. It is given by
\begin{align} 
\vectwo{E'}{B'}= \mattwo{\cos{(\omega \overline{d} /c)}}
{i n^{-1} \sin{((\omega \overline{d} /c))}} {i n \sin{((\omega
\overline{d} /c))}} {\cos{(\omega \overline{d} /c)}} \vectwo{E}{B}
\label{transfer} 
\end{align} 
where $\omega$ is the angular frequency, $\overline{d}=nd$ is the
propagation length of the layer and $c$ is the speed of light in
vacuum. Multiplying matrices for individual layers gives the transfer
matrix for a structure, which can be used to calculate spectral
properties such as transmission. The spectral symmetries of the
transfer matrix follow directly from the periodicity and symmetries of
the trigonometric functions in it.  Apart from trivial phases, it is
periodic in frequency, with period $\pi c/ \overline d$, and symmetric
about values of $\omega$ which are integer multiples of $\pi c/(2
\overline{d})$. If $\overline{d}$ is the same for every layer, the
transfer matrix for the whole structure will
display these symmetries. 

We next demonstrate that such chiral multilayers form exact
implementations of an SSH chain. Consider a structure where each
layer has the same propagation length, $\overline{d}$. Defining
the fields $E_i$ and $B_i$ at the $i^{\rm th}$ interface,
the transfer matrices can be used to eliminate the $B_i$, giving a
relationship between the $E_i$:
\begin{align}
n_{i-1,i} \, E_{i-1}+n_{i,i+1} \, E_{i+1}=
 (n_{i-1,i}+n_{i,i+1}) \, {\cos{(\omega \overline{d}/c)}} \, E_i 
 \label{tb1}
\end{align}
Here, $n_{i,i+1}$ is the refractive index of the layer between the 
$i^{\rm th}$ and $(i+1)^{\rm th}$ interface. If we now define an `energy'
$\varepsilon=\cos{(\omega \overline{d}/c)}$ and scaled fields 
$\tilde{E}_i=(n_{i-1,i}+n_{i,i+1})^{1/2}E_i$, this becomes a tight-binding
model,
\begin{align}
 t_{i-1,i} \, \tilde{E}_{i-1} +  t_{i,i+1} \, \tilde{E}_{i+1} 
 = \varepsilon \tilde{E}_i,
\end{align}
where the hopping amplitude is
\begin{align}
t_{i,i+1}=\frac{n_{i,i+1}}{[(n_{i-1,i}+n_{i,i+1})(n_{i,i+1}+n_{i+1,i+2})]^{1/2}}.
\label{hop}
\end{align} 
This tight binding model has the form of an SSH chain: there are only
nearest-neighbour hoppings and the on-site energies are all zero.

When applied to the case of a conventional bilayer Bragg stack,
Eq.(\ref{hop}) gives two values for the hopping amplitudes,
$n_1/(n_1+n_2)$ and $n_2/(n_1+n_2)$, for the layers with index $n_1$
and $n_2$, respectively. The tight binding system is then identical to
the simplest SSH model.  If the layer propagation lengths are
commensurate rather than identical, it is necessary to divide the
layers into sub-units of equal thickness. The tight binding model then
has more than two layers per unit cell, corresponding to generalised
SSH systems, such as the SSH-4 model\cite{ssh4-maffei, ssh4-xie}.

A conventional microcavity structure is made by joining two Bragg
stacks with the same termination, giving a defect with twice the
thickness of a normal layer. The doubled layer corresponds to two hops
with the same amplitude, so this is an exact experimental realisation
of the interface between SSH domains of different topology. The cavity
mode is found at the mid-gap chiral point, and corresponds to the
topologically protected interface state. This means that its frequency
should be unaffected by disorder consisting of variations in the
refractive indicies of the layers of the structure. However, chiral
symmetry needs to be maintained, so the propagation length,
$\overline{d}=nd$, of each layer must be kept constant, by
adjusting its thickness when the refractive
index is changed.

\begin{figure}

\begin{center} \mbox{ \includegraphics[scale=0.22]{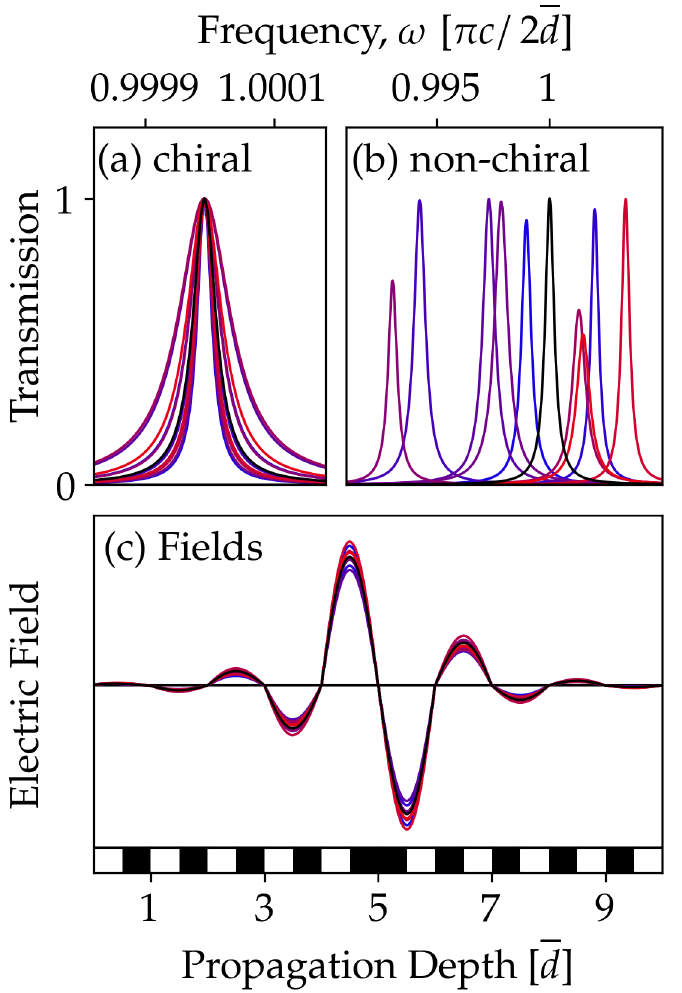} } \end{center}
\caption{Topological protection for a chiral multilayer photonic structure. 
(a) and (b) show calculated transmission spectra, in the frequency range
around the middle of the first band-gap, for a cavities created by joining two
disordered Bragg mirrors. Each mirror consists of five periods, terminated at the
interface with a high index layer. The structures are designed (see text) 
so that both have a mid-gap cavity mode; in (a) this corresponds
to a chiral point, making the mode topologically protected, while in (b) the
chiral point is in a higher gap, so there is no protection. 
The colours correspond to ten instances of
disorder, in which the refractive index in each layer is varied randomly by
$\pm 20\%$, from the nominal values of $n_1=3$ and $n_2=1$, while the layer
thickness is adjusted to keep the propagation length $\overline{d}=nd$ constant.
The black curves are reference spectra without disorder.
(c) shows electric field profiles for the protected
mode in (a), plotted as a function of the optical depth, so the layer interfaces,
shown in the black and white bar, coincide for all structures.
} 
\label{fig-cavity}
\end{figure}

In Fig.\ref{fig-cavity}, we show transmission spectra, calculated
using the transfer matrix method, which demonstrate topological
protection for a microcavity in which the refractive index of each
layer is varied randomly, adjusting the thickness to maintain the
chiral condition. We consider two structures, with and without a
chiral symmetry point in the lowest gap. The first
(Fig.\ref{fig-cavity}(a)) is a conventional cavity satisfying the
Bragg condition $n_1 d_1=n_2 d_2$, giving a chiral point in the middle
of the gap, so the cavity mode is topologically protected. The second
structure (Fig.\ref{fig-cavity}(b)) is more complicated. The
transmission thickness of a period, $n_1 d_1 + n_2 d_2$ is constant,
but the ratio of high to low index material is reversed at the
interface: on one side $n_1 d_1=3 n_2 d_2$, while on the other $n_2
d_2 = 3 n_1 d_1$. These mirrors have identical bandstructures and the
interface supports a cavity mode in the middle of every gap.
However, the structure maps onto an SSH-4 model, so there is no chiral
point in the first gap, and no topological protection.  The
difference is evident in the spectra: for the chiral case, there is
absolutely no change in the energy of the cavity peak, while without
the protection, each instance of the disorder gives a distinct peak.
In Fig.\ref{fig-cavity}(c), we plot the electic field profile of the
mode for the chiral case.  There is a different profile for each
instance of the disorder, but in all cases the field goes to zero at
alternate interfaces.  This relates to the form of the protected state
in SSH systems, which has zero amplitude on alternate sites.

The main factor determining how closely exact chiral symmetry can be
approached is likely to be the accuracy of the fabrication, ensuring
that the propagation length of each layer is the same. The
description in terms of a constant, real refractive index also needs
to be considered. The frequency dispersion of the index is not really
an issue, because the topological protection only requires the
matching to be exact at the chiral point.
Absorption, corresponding to a complex refractive index $n+i \kappa$,
can break the chiral symmetry. However, it produces fractional
energy shifts $\sim (\kappa/n)^2$, which can be made very small with
appropriate material choices. Furthermore, a mode of quality factor
$Q$, without absorption, will also be weakened to the point where it
becomes unobservable when $\kappa/n \sim 1/Q$, so the
maximum detectable shift due to absorption will be of order of the
linewidth divided by $Q$. $Q$ values of 
thousands are achievable in semiconductor microcavities, so this is a
small effect.

\section*{Transmission Line Structures}

The chiral and topological properties discussed above follow from the
form of the transfer matrices which describe the structures. Similar
mathematics occurs in other situations where waves are
one-dimensional, such as acoustic waves in multilayers and waves on
strings. Hence, it should be possible to find the physics of
topological protection in these systems. We shall consider the
propagation of radio-frequency signals in coaxial cable transmission lines.
For an ideal, lossless, cable, the transfer matrix relates the voltage, $V$, and
current, $I$ at either end, according to 
\begin{align}
\vectwo{V'}{I'}= \mattwo{\cos{(\omega {d}/v)}}{i {Z}\sin{(\omega
{d}/v)}} {{i}{Z^{-1}}\sin{(\omega {d}/v)}}{\cos{(\omega {d}/v)}}
\vectwo{V}{I}, 
\end{align} where the impedance
$Z=\sqrt{L/C}$, and the propagation velocity $v=1/ \sqrt{LC}$, with
$L$ and $C$ the inductance and capacitance per unit length. This has
the same form as Eq.(\ref{transfer}), so exact
analogues of photonic multilayer structures can be constructed by
joining sections of transmssion line with different
impedances\cite{coax-schneider,coax-munday,coax-sanchez-lopez}.

\begin{figure}

\begin{center} \mbox{ \includegraphics[scale=0.27]{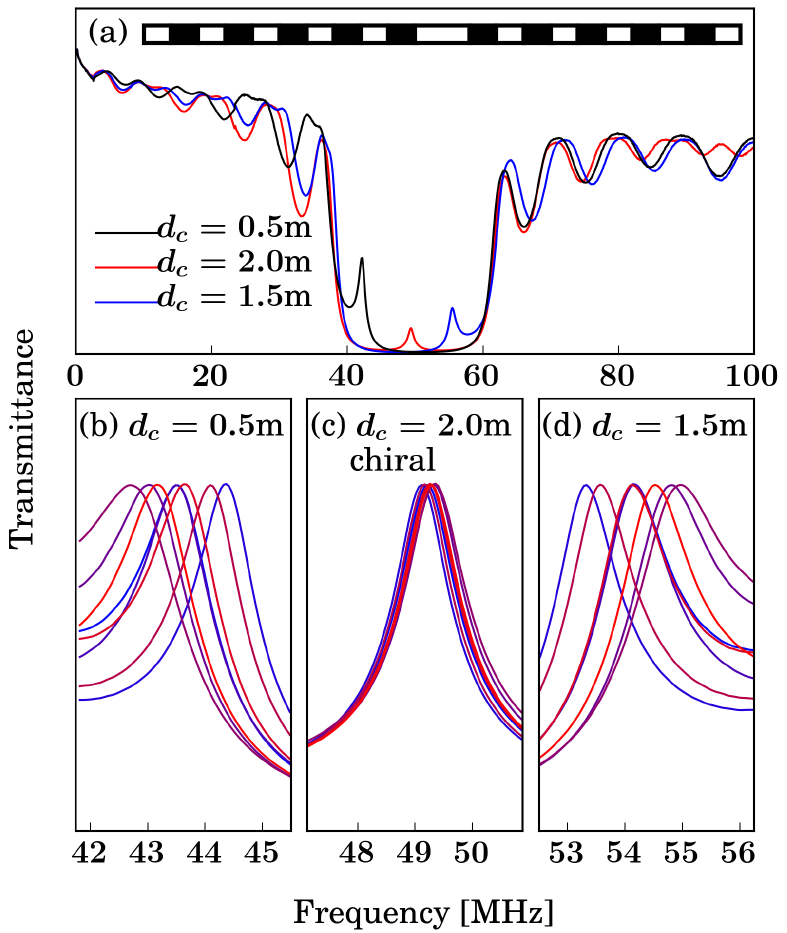} } \end{center}
\caption{ Experimental transmittance spectra for cavity structures made from coaxial
  cables. (a) spectra for structures without 
  disorder, shown schematically in the inset. Each has two chiral
  mirrors, made from five repeats of a pair of cables with impedances
  $50 \Omega$ (white) and $93 \Omega$ (black), surrounding a $50
  \Omega$ section defining the cavity.  The physical lengths of the
  $50 \Omega$ and $93 \Omega$ cables in the mirrors are 1.0m and 1.24m
  respectively, giving the same propagation lengths. Three 
  cavity lengths, $d_c$, are used: for $d_c=2.0$m, the propagation
  length is double that of the mirror layers, so the chiral symmetry
  is maintained, while for $d_c=0.5$m and 1.5m, it is broken. The
  band-gap corresponds to the strong dip in transmission, between
  about 40 and 60 MHz, containing the peak due to the
  cavity mode. (b)-(d) show normalised spectra, in the region of this 
  mode, for eight different structures with disorder, where half of
  the $93 \Omega$ sections are selected randomly and replaced with $75
  \Omega$ cables.  Identical mirrors are used with each
  cavity, indicated by the line colours. In the
  chiral case, (c), the disorder has very little effect on the mode
  frequency, demonstrating that it is topologically protected} 
\label{fig-coax}
\end{figure}

Transmission lines made from coaxial cable sections with the same
propagation length $\overline{d}=d c/v$, are a convenient system for
demonstrating the topological protection discussed above. It is
straightforward to make an ensemble of instances of disorder by
swapping in sections of cable with different impedances. Transmittance
spectra for cavities constructed in this way are shown in Fig.\ref{fig-coax}.
We consider three cavity structures: in each case, the mirrors
consist of a repeated chiral bilayer, but in one the chiral symmetry
is maintained as the cavity is formed by  repeating one of the low
impedance layers ($d_c=2.0$m), while the other two deliberately break
the symmetry, with cavities of different lengths. When disorder is
introduced into the mirrors, the mode in the chiral
cavity undergoes very little change in frequency, demonstrating the
topological protection. By contrast, the
non-chiral structures show a broad spread in frequency between the
different instances of the disorder. However, even in the non-chiral
structures, the mirrors on either side of the cavity layer retain
their chiral symmetry. This accounts for the near reflection symmetry
between equivalent spectra in Fig.\ref{fig-coax}(b) and (d): the
cavity modes are approximately equidistant in frequency from the
chiral point, so they experience similar changes in mirror
reflectivity for a given instance of the disorder.

With transmission lines, we can go further than simple one-dimensional
chains, since more than two cables can meet at a junction. We then
have a network rather than a simple transmssion line. When the
propagation lengths, $\overline{d}$, of all the cables is the same,
our result, Eq.(\ref{tb1}), generalises to \begin{align} \sum_j
Z_{ij}^{-1} \, V_j = \sum_j Z_{ij}^{-1} \cos{(\omega \overline{d}/c)}
\, V_i \end{align} where the sums are over the nearest neighbours of
site $i$, to which it is directly connected. A similar result has been
derived by Zhang and Sheng\cite{network-zhang}, for microwave
waveguide networks.  With the equivalent rescaling and identification
of $\varepsilon=\cos{(\omega \overline{d}/c)}$, this again looks like
a tight binding model. Provided any loops have an even number of
nodes, this network will have chiral symmetry, and so potentially
non-trivial topology.  In such a network the spatial positioning of
sites is unimportant, the only relevant consideration being how they
are connected. Hence lattices can be fabricated which are impossible
under the physical constraints of three dimensional
space\cite{topology-ryu,hall4d-zhang} and Euclidean
geometry\cite{hyperbolic-kollar}.  Transmission line structures can
also be seen as a bridge between photonic systems and topological
circuits\cite{electrical-ningyuan, electrical-lee}. It is natural to
combine transmission lines with discrete electronics located at the
nodes, providing, for example, gain and loss or nonlinearity. This
would extend the range of models which can be studied to include PT
symmetric\cite{PT-ganainy} and nonlinear systems.

To conclude, we have shown that commensurate photonic multilayers
provide exact implementations of SSH chains, with perfect chiral
symmetry. A Bragg stack, with equal propagation lengths for the layers,
corresponds to the basic SSH model, and conventional microcavity
modes are  the interface states between regions of
different topology. We have demonstrated this experimentally using
coaxial transmission lines, a mathematically analogous system,
where we show that the cavity mode is topologically protected against
a particular class of disorder.  Finally, we have shown that the same
mathematics works in transmission line networks, suggesting that they
are a promising platform for more general studies of topological
physics.

\begin{acknowledgments}
We wish to thank E. Chekhovich for help with setting up the experiment.
\end{acknowledgments}


\begin{thebibliography}{10}

\bibitem{SSH}
W. P. Su, J.~R.~Schrieffer and A. J. Heeger,
Phys. Rev. Lett. {\bf 42}, 1698 (1979).

\bibitem{photonic-ozawa}
T.~Ozawa, H.~M.~Price, A.~Amo, N.~Goldman, M.~Hafezi, L.~Lu, M.~C.~Rechtsman, D.~Schuster, 
J.~Simon, O.~Zilberberg and I.~Carusotto,
Rev. Mod. Phys. {\bf 91,} 015006 (2019).

\bibitem{photonic-malkova}
N.~Malkova, I.~Hromada, X.~Wang, G.~Bryant  and Z.~Chen, 
Opt. Lett. {\bf 34}, 1633 (2009).

\bibitem{microwave-poli}
C.~Poli, M.~Bellec, U.~Kuhl,  F.~Mortessagne and H.~Schomerus, 
Nat. Commun. {\bf 6}, 6710 (2015).

\bibitem{polariton-st-jean}
P.~St-Jean, V.~Goblot,   E.~Galopin,   A.~Lema\^itre,
T.~Ozawa, L.~Le~Gratiet, I.~Sagnes,  J.~Bloch and  A.~Amo 
Nat. Photonics {\bf 11}, 651 (2017)

\bibitem{plasmon-bleckmann}
F.~Bleckmann, Z.~Cherpakova, S.~Linden and A.~Alberti, 
Phys. Rev. B{\bf 96}, 045417 (2017).

\bibitem{acoustic-li}
X~Li, Y.~Meng, X.~Wu, S.~Yan, Y.~Huang, S.~Wang and W.~Wen
Appl. Phys. Lett. {\bf 113}, 203501 (2018).

\bibitem{acoustic-esmann}
M.~Esmann, F.~R.~Lamberti, A.~Lema\^itre and N.~D.~Lanzillotti-Kimura,
Phys. Rev. B {\bf 98}, 161109(R) (2018).


\bibitem{atom-atala}
M.~Atala,  M.~Aidelsburger  J.~T.~Barreiro,  D.~Abanin,
T.~Kitagawa,  E.~Demler    and T.~Bloch
Nat. Phys. {\bf 9}, 795-800 (2013).



\bibitem{microcavity-weisbuch}
C.~Weisbuch, M.~Nishioka, A.~Ishikawa and Y.~Arakawa,
Phys. Rev. Lett. {\bf 69}, 3314 (1992).

\bibitem{microcavity-skolnick}
M.~S.~Skolnick, T.~A.~Fisher and D.~M.~Whittaker,
Semicond. Sci. Technol. {\bf 13}, 645-669 (1998).

\bibitem{inversion-zak}
J.~Zak, Phys. Rev. Lett. {\bf 62} 2747 (1989).

\bibitem{multilayer-xiao}
Meng~Xiao, Z.~Q.~Zhang and C.~T.~Chan,
Phys. Rev. X {\bf 4}, 021017 (2014).


\bibitem{coax-schneider} 
G.~J.~Schneider, S.~Hanna, J.~L.~Davis and G.~H.~Watson,
J. Appl. Phys. {\bf 90}, 2642 (2001).

\bibitem{coax-munday}
J.~N.~Munday and W.~M.~Robertson, 
Appl. Phys. Lett. {\bf 83}, 1053 (2003).

\bibitem{coax-sanchez-lopez}
M.~M.~S\'anchez-L\'opez, J.~A.~Davis and K.~Crabtree, 
Am. J. Phys. {\bf 71}, 1314 (2003).


\bibitem{ssh4-maffei} M.~Maffei, A.~Dauphin, F.~Cardano, M.~Lewenstein and
 P.~Massignan,
New J. Phys. {\bf 20} 013023 (2018).

\bibitem{ssh4-xie} D.~Xie, W.~Gou, T.~Xiao, B.~Gadway and B.~Yan, 
NPJ Quantum Inf. {\bf 5}, 55 (2019). 


\bibitem{network-zhang}
Z.~Q.~Zhang and P. Sheng,
Phys. Rev. B {\bf 49}, 83 (1994).

\bibitem{topology-ryu} S.~Ryu, A.~P.~Schnyder, A.~Furusaki and A.~W.~W.~Ludwig,
New J. Phys. {\bf 12}, 065010 (2010).

\bibitem{hall4d-zhang}
S.-C.~Zhang and J.~Hu, Science {\bf 294}, 823 (2001).

\bibitem{hyperbolic-kollar}
A.~J.~Koll\'ar, M.~Fitzpatrick and A.~A.~Houck,
Nature {\bf 571}, 45 (2019).



\bibitem{electrical-ningyuan}
J. Ningyuan, C.~Owens, A.~Sommer and D.~Schuster and J.~Simon
Phys. Rev. X, {\bf 5}, 021031 (2015).


\bibitem{electrical-lee}
C.~H.~Lee, S.~Imhof, C.~Berger, F.~Bayer, J.~Brehm, L.~W.~Molenkamp,
T.~Kiessling and R.~Thomale, 
Commun. Phys. {\bf 1} 39 (2018).


\bibitem{PT-ganainy}
R.~El-Ganainy,  K.~G.~Makris,  M.~Khajavikhan,  Z.~H.~Musslimani,
S.~Rotter  and D.~N.~Christodoulides, 
Nat. Phys. {\bf 14} 11 (2018).



\end{thebibliography}
\end{document}